\documentclass[prl,aps,twocolumn,showpacs,preprintnumbers,superscriptaddress]{revtex4}
\usepackage{graphicx}
\usepackage{dcolumn}
\usepackage{bm}
\usepackage{amssymb}
\usepackage{pifont}
\usepackage{amsmath}
\usepackage{times}
\usepackage[nooneline]{subfigure}

\begin{document}

\title{Time dependent couplings and crossover length scales
in non-equilibrium surface roughening}

\author{Marc Pradas}
\affiliation{Departament d'Estructura i Constituents de la Mat{\`e}ria, 
Universitat de Barcelona, Av. Diagonal 647, E-08028 Barcelona, Spain}

\author{Juan M.\ L{\'o}pez}\email{lopez@ifca.unican.es}
\affiliation{Instituto de F{\'i}sica de Cantabria (IFCA),
CSIC--UC, E-39005 Santander, Spain}

\author{A.\ Hern{\'a}ndez-Machado}
\affiliation{Departament d'Estructura i Constituents de la Mat{\`e}ria, 
Universitat de Barcelona, Av. Diagonal 647, E-08028 Barcelona, Spain}

\date{\today}

\begin{abstract}
We show that time dependent couplings may lead to 
nontrivial scaling properties of the surface fluctuations 
of the asymptotic regime in non-equilibrium kinetic roughening models .
Three typical situations are studied.
In the case of a crossover between two different rough regimes, 
the time-dependent coupling may result in anomalous scaling for 
scales above the crossover length. In a different setting, for a 
crossover from a rough to either a flat or damping regime, the time 
dependent crossover length may conspire to produce a
rough surface, despite the most relevant term tends to 
flatten the surface. In addition, our analysis sheds light into an 
existing debate in the problem of spontaneous imbibition, 
where time dependent couplings naturally arise in theoretical 
models and experiments.
\end{abstract}


\maketitle

The theory of surface growth has applications to phenomena
that cover a wide range of length scales from nanometers to millimeters,
including the growth of thin films from an incoming flux of atoms
\cite{barabasi,krug,pimpi},
fluid imbibition in porous media \cite{alava}, and propagation of 
fracture cracks
\cite{bouchaud}, among many others. It is often 
observed that surfaces kinetically roughen and become scale-invariant.

In kinetic roughening details of the interactions are largely
irrelevant in the mathematical description of the critical properties
of the surface at long wavelengths, akin to critical point phenomena.
Therefore, in the hydrodynamic limit, one can
describe surface growth in $d+1$ dimensions by the stochastic equation
\begin{equation}
\label{langevin} \frac{\partial{h}}{\partial{t}} =
{\cal G}(\nabla h) + \eta(\mathbf{x},t),
\end{equation}
where $h({\mathbf{x}},t)$ is the height of the interface at substrate
position ${\mathbf{x}}$ and time $t$. The functional ${\cal G}(\nabla h)$
depends on the specific model and should satisfy  
all the symmetries and conservation laws. The external 
noise $\eta({\mathbf{x}},t)$ 
describes the random driving forces acting on the surface, for instance 
the influx of particles in a deposition processes.

We shall be discussing here surface growth models with local coupling
among degrees of freedom, {\it i.e}, growth equations that include
only terms that depend on derivatives of the height, $\Upsilon=\nabla h$.
In the spirit of Ginzburg-Landau-Wilson
theory of critical phenomena the functional ${\cal G}(\Upsilon)$
is constructed as the
leading-order expansion in powers of the argument $\Upsilon$, its derivatives,
and combinations thereof. One explicitly avoids to
include terms that are incompatible with the symmetries of the problem.
The corresponding expansion takes the form of a sum,
${\cal G}(\Upsilon) = \sum_{i} \omega_i \Phi_i(\mathbf{x},t)$,
where $\omega_i$ are the {\em coupling constants} and $\Phi_i(\mathbf{x},t)$
are the {\em local operators} in the terminology of the
renormalization group (RG).
Local operators correspond to the combinations of degrees of freedom and
its derivatives, which typically include surface diffusion $\nabla^2 h$,
curvature diffusion $\nabla^4 h$, and in general higher-order diffusion terms
$\nabla^{2m} h$. Also nonlinear terms usually appear, $(\nabla h)^{2n}$,
$\nabla^{2n} (\nabla h)^{2m}$, and so on. The asymptotic long wavelength
limit is governed by the most relevant terms in the RG sense.
Higher-order terms are nonetheless important in
describing crossover effects before the truly
asymptotic behavior is reached. This leads to the existence of crossover
length scales and characteristic times at which
one can observe the true asymptotic scaling behavior of the system.
In the case of time independent coupling parameters
the analysis of crossover length scales is relatively simple and very
well studied \cite{barabasi,krug,pimpi}. However, little is known about
less common situations in which the couplings depend explicitly on time.
Examples include the problem of
a stable phase growing at the expense of a metastable phase \cite{emilio},
and spontaneous imbibition \cite{alava}, among others. Indeed, 
the problem of spontaneous imbibition has been a subject of great interest in 
the last few years \cite{alava,dube,aurora,paune,marc,laurila}. 
Different theoretical 
approaches have arrived at the conclusion that, for small 
deviations around of the mean position $H(t)$, the fluid-fluid 
interface (in Fourier space) is given by
\begin{equation}
\label{eq_imbi}
\frac{\partial \widehat{h}_k(t)}{\partial t} = -\sigma K |k|k^2 \widehat{h}_k - 
\dot H(t) |k| \widehat{h}_k + K |k| \widehat{\eta}_k,
\end{equation}
where $\sigma$ and $K$ are the surface tension and permeability constants, 
respectively, and the average position follows Washburn's law 
$H(t) = \sqrt{H_0^2 + 2at}$ and arises from mass conservation. There is an
interesting debate regarding the role of this time-dependent coupling in the
scaling observed in both, simulations \cite{dube,marc,laurila} 
and experiments \cite{soriano,soriano2,gero}. 
This problem has largely motivated our study of the scaling 
of the surface fluctuations in systems where couplings 
depend explicitly on time.

In this Letter we study the interplay between
a crossover length scale growing in time and 
the dynamic correlation length characterizing the kinetic 
roughening process. We find nontrivial scaling properties, 
including anomalous roughening \cite{lopez97,lopez99}.
The long time limit scaling behavior strongly depends on 
the nature of the phases that the dynamic
crossover separates. We focus on three typical cases that cover the most
important situations one can find: (i) Crossover between two different
rough regimes, (ii) crossover from a rough to a flat regime, and
(iii) the existence a damping term, where scale-invariant
fluctuations are damped over a certain length scale that varies with time.
We exemplify our general scaling analysis 
with simple model systems in which dimensional
analysis gives the exact exponents. In order to gain analytical 
understanding we restrict ourselves to linear model examples that allow 
exact computation of the critical exponents. Results are compared with 
numerical simulations.

\paragraph{Crossover between two different rough regimes.-}
Consider a growing surface described by Eq.\
(\ref{langevin}). Invariance under translation
along the growth and substrate directions as well as invariance in
the election of the time origin rule out an explicit dependence of
${\cal G}$ on $h$, ${\mathbf{x}}$ and $t$. These symmetry requirements
alone lead to scale invariant growth in a generic fashion \cite{barabasi}.
Let us consider the two most relevant terms in 
the leading-order expansion of ${\cal G}(\nabla h)$ 
in the hydrodynamic limit, so that we have ${\cal G} =
\omega_\mathrm{I} \Phi_\mathrm{I}(x,t) + \omega_\mathrm{II}
\Phi_\mathrm{II}(x,t) + $ higher-order terms. Let the $\Phi_\mathrm{I}$
term be more relevant (in the RG sense) than $\Phi_\mathrm{II}$. We first 
discuss the case in which Eq.\ (\ref{langevin}) exhibits crossover
between two different rough regimes. This means $\omega_\mathrm{II}
\Phi_\mathrm{II}(x,t)$ is relevant at short scales, while $\omega_\mathrm{I}
\Phi_\mathrm{I}(x,t)$ becomes the most dominant in the long wavelength
limit. Hence one expects to find an early times 
(short scales) regime with scaling
exponents $\alpha^\mathrm{(II)}$ and $z^\mathrm{(II)}$ that crosses over to the
true asymptotic regime, governed by the operator $\Phi_\mathrm{I}$, with
exponents $\alpha^\mathrm{(I)}$ and $z^\mathrm{(I)}$. Dimensional
analysis indicates that there exits a crossover length $\ell_{\times} \sim
(\omega_\mathrm{II}/\omega_\mathrm{I})^{1/q}$, for some exponent $q$. 
This is the typical length above which the most relevant
term $\omega_\mathrm{I} \Phi_\mathrm{I}$ takes over. This is likely
the most common situation of crossover behavior in surface growth. 

The question we wish to address here is how this picture is modified when the
most relevant operator's coupling, $\omega_\mathrm{I}$, decreases in time. The
most interesting situation naturally arises for a power-law decay of the
coupling $\omega_\mathrm{I}(t) \sim t^{-\gamma}$, so that the term
$\omega_\mathrm{I}(t) \Phi_\mathrm{I}$ gets effectively less relevant for
longer times as compared with the $\omega_\mathrm{II} \Phi_\mathrm{II}$ term.
In this case the crossover length diverges in time as a power-law,
$\ell_{\times}(t) \sim t^{\gamma/q}$, and the interplay with the other relevant
scale in the problem --namely, correlation length $\xi(t) \sim t^{1/z}$, will
give rise to an interesting behavior. On scales smaller than $\ell_{\times}(t)$
the less relevant term II dominates, while we expect a crossover to the 
asymptotic regime I for $\xi(t) \gg \ell_{\times}(t)$. The caveat is now that 
the observable scaling regimes crucially depend on the exponent of the 
coupling $\gamma$. If $\gamma < q/z^\mathrm{(II)}$ the crossover does
take place as described above. On the contrary, 
for $\gamma > q/z^\mathrm{(II)}$ the crossover length grows at an
exponentially faster rate than the correlation length. Therefore, 
in this case the time-dependent coupling
effectively takes the crossover length to scales much larger than those  
that can be correlated by the dynamics at any finite time. As a consequence the
system will never cross over to regime I, and only the "less" 
relevant operator II governs 
the scaling regime observable for arbitrarily large system sizes. 

We now study a simple $1+1$ dimensional model that exemplifies 
the crossover behavior
between two different scaling regimes above discussed. 
We consider (\ref{langevin}) with
$\Phi_\mathrm{I} = \nabla^2 h$ and $\Phi_\mathrm{II} = - \nabla^4h$, 
describing diffusive coupling. Then we have the growth equation
\begin{equation}\label{eq:int}
\frac{\partial{h}}{\partial{t}} = \nu(t)\nabla^{2}h - \nabla^{4}h + \eta(x,t),
\end{equation}
where we have rescaled all the couplings but $\nu(t)$ to unity. The noise is
delta-correlated, $\langle\eta(x,t)\eta(x',t')\rangle= 2
\delta(x-x')\delta(t-t')$. Higher-order diffusion terms, $\nabla^{2n}h$, might
be present, but are irrelevant for the scaling behavior. Actually, even the
$\nabla^4 h$ term is irrelevant as compared with $\nabla^2h$ if the couplings
are independent of time. Consider now a decaying coupling $\nu(t) \sim
t^{-\gamma}$ \cite{note}. Balancing the two gradient terms in Eq.\
(\ref{eq:int}) one finds that they become comparable at the typical crossover 
scale $\ell_{\times}(t) \sim t^{\gamma/2}$, with the exponent 
$q=2$ in our previous analysis.

Dimensional analysis tells us that the kinetic roughening process is governed 
by the fourth derivative term (that
we label as II) on scales smaller than $\ell_{\times}$, while the diffusion
term should dominate on much larger scales (that we label as regime I). 
Below the crossover, for $\xi(t) \ll \ell_{\times}(t)$, 
Eq.\ (\ref{eq:int}) has a
correlation length $\xi(t) \sim t^{1/z^\mathrm{(II)}}$, where $z^\mathrm{(II)}
= 4$, given by the $\nabla^4 h$ term dynamics. Crossover to the asymptotic
regime I takes place when the correlation length reaches the crossover length
$t^{1/z^\mathrm{(II)}} \sim t^{\gamma/2}$. Following our previous analysis 
we expect that the regime I can only be reached if 
$\gamma < 2/z^\mathrm{(II)}$, {\it i.e},
$\gamma < 1/2$. On the contrary, for $\gamma > 1/2$ the crossover 
is wiped out  and the less relevant operator II (in RG sense)
does govern the scaling in the long wavelengths limit as well.
This analysis is in excellent agreement with simulations as 
shown in Figure 1.

\begin{figure}
 \includegraphics[width=0.40\textwidth]{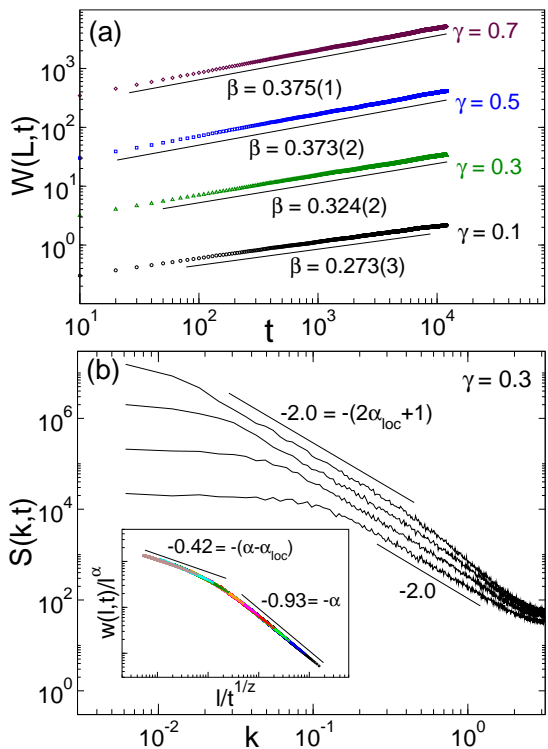}
\caption{(Color online) Crossover from rough to rough. 
Numerical integration of Eq. (\ref{eq:int}) 
using a time step $\Delta t=10^{-2}$ and $c=10^{-3}$ in a system size 
$L=1024$. All results were averaged over 200 realizations. 
(a) Global width of the interface, 
computed for different $\gamma$ values. The theoretical prediction for the 
growth exponent $\beta^\mathrm{(I)} =(1+\gamma)/4$ for $\gamma < 1/2$
fully agrees with simulations.
For any $\gamma > 1/2$ we obtain  
$\beta=3/8$, as expected. The curves are vertically shifted for clarity. 
(b) Power spectrum for 
$\gamma=0.3$ evaluated at times $10$, $10^{2}$, $10^{3}$, and $10^{4}$. 
Anomalous scaling with a local 
roughness exponent $\alpha^\mathrm{(I)}_\mathrm{loc}=0.5$ is observed. The inset
shows a data collapse of the local width data for
$\alpha^\mathrm{(I)}=0.93(1)$ and $z^\mathrm{(I)}=2.86(4)$ to
be compared with the predicted values 
$\alpha^\mathrm{(I)}=0.928$ and $z^\mathrm{(I)}=2.857$ for $\gamma = 0.3$.\label{fig1}}
\end{figure}

Let us now study the scaling properties in the case $\gamma <
1/2$, where one expects two different regimes with different
critical exponents. Smaller
scales, regime II, are simply governed by the dynamics of the curvature
diffusion term $\nabla^4 h$ and we find the well-known results
$z^\mathrm{(II)}=4$, $\alpha^\mathrm{(II)}=3/2$, and the local roughness
exponent $\alpha^\mathrm{(II)}_\mathrm{loc}=1$ \cite{lopez97,lopez99}.
Above the crossover, $\xi(t) \gg \ell_{\times}$, the system is
expected to be in regime I and a scaling analysis gives the exact global
exponents $\alpha^\mathrm{(I)} = (1+\gamma)/[2(1-\gamma)]$, 
$z^\mathrm{(I)} = 2/(1-\gamma)$, $\beta^\mathrm{(I)} =
(1+\gamma)/4$, 
and the scaling relation $\alpha^\mathrm{(I)} = z^\mathrm{(I)}
\beta^\mathrm{(I)}$ is fulfilled. Figure \ref{fig1} shows that this  
scaling analysis is in excellent agreement with a 
numerical integration of the model (\ref{eq:int}).

However, even for $\gamma < 1/2$ the asymptotic scaling behavior is 
nontrivial, since the time-dependent coupling $\nu(t)$
leads to anomalous scaling of the local surface fluctuations. This is
more clearly proved in Fourier space in terms of the
structure factor (or spectral power spectrum), $S(k,t)=\langle
\widehat{h}(k,t) \widehat{h}(-k,t)\rangle$, where
$\widehat{h}(k,t)$ is the Fourier transform of the surface in a system
of lateral size $L$, $\widehat{h}(k,t) = L^{-1/2} \sum_x
[h(k,t) - \overline h(t)] \exp(ik x)$.
>From the growth equation (\ref{eq:int}) and neglecting the $\nabla^4 h$ term
we obtain
\begin{equation}\label{eq:SPC}
S(k,t)\sim k^{-2}t^{\gamma}s(k^{2}t^{1-\gamma}),
\end{equation}
where the scaling function $s(u)$ has the asymptotes, $s(u) \sim u$ for $u \ll 1$, and
$s(u)\sim \mathrm{constant}$ for $u \gg 1$~\cite{note2}. The temporal shift in the
power spectrum, $t^\gamma$, implies anomalous scaling of the local fluctuations. The
theory of anomalous scaling \cite{lopez97,lopez99} tells us that the scaling relation
$\gamma=2(\alpha^\mathrm{(I)}-\alpha^\mathrm{(I)}_\mathrm{loc})/z^\mathrm{(I)}$ must be
fulfilled. We then find that, according to the spectral density (\ref{eq:SPC}), the
scaling in the asymptotic regime (regime I) is anomalous in this case with a local
roughness exponent $\alpha^\mathrm{(I)}_\mathrm{loc} = 1/2$. Note that the local roughness
exponent does not depend on the coupling exponent $\gamma$ and self-affinity,
$\alpha^\mathrm{(I)}=\alpha^\mathrm{(I)}_\mathrm{loc}=1/2$, is recovered whenever the
coupling is time independent ($\gamma = 0$). A comparison with numerical simulations for
$\gamma=0.3$ is shown in Figure \ref{fig1}.

\paragraph{Crossover from a rough to a flat regime.-}
Another interesting situation 
occurs when the most relevant operator happens to lead to a flat surface. In 
this case the crossover is expected to take 
place from an early times (short scales) rough 
regime II, with roughness exponent $\alpha^\mathrm{(II)} > 0$ 
to an asymptotic flat regime, where $\alpha^\mathrm{(I)} = 0$. Again, a time 
dependent crossover length leads to nontrivial scaling properties in this case.
Below the crossover we expect a rough 
phase dominated by the $\Phi_\mathrm{II}$ term with exponents 
$\alpha^\mathrm{(II)}>0$ and $z^\mathrm{(II)}$.
Above the crossover, however, surface correlations 
cannot evolve any longer since the operator $\Phi_\mathrm{I}$ does not 
amplify surface fluctuations (note that
$\alpha^\mathrm{(I)} = 0$ and $z^\mathrm{(I)}=0$). This leads to an 
asymptotic roughness exponent 
identical to that 
in the early regime. 
Moreover, it is worth to stress here that the dynamical length scale
$\sim t^{1/z^\mathrm{(I)}}$ 
becomes a constant above the crossover (note that 
$z^\mathrm{(I)} = 0$).
Therefore, above the crossover 
the only relevant scale left in the problem scales as $\sim t^{\gamma/q}$. 
This must be identified with the correlation length $\xi(t)$ above 
the crossover. We therefore conclude that the time-dependent 
coupling leads to an asymptotic regime with exponents 
$\alpha=\alpha^\mathrm{(II)}$
and $z = q/\gamma$, instead of those that would naively arise 
from the operator I, $\alpha^\mathrm{(I)}=z^\mathrm{(I)}=0$. 
It is remarkable that, despite the most relevant term
tends to flatten the surface, the time-dependent coupling conspires to 
produce a rough surface. The roughness exponent is inherited from the early
time (short scale) phase. Meanwhile, the time-dependent crossover 
length becomes the only relevant scale above the crossover, 
which is to be associated with the dynamical
exponent $z$ of the system in the long times regime.

To exemplify this general scaling analysis 
we now study a simple $1+1$ dimensional model system exhibiting a crossover
between rough regime on short scales to a flat surface regime in the 
long wavelengths limit.
Let us consider same growth model as in (\ref{eq:int}),
but the noise is now conserved, $\langle\eta_{c}(x,t)\eta_{c}(x',t')\rangle
= -2\nabla^{2}\delta(x-x')\delta(t-t')$. Note that the surface diffusion 
operator $\nabla^2h$ leads to flat surface fluctuations, 
$\alpha^\mathrm{(I)}=z^\mathrm{(I)}=0$, in the presence 
of conserved noise.

On small scales the surface is
expected to be dominated by the less relevant $\nabla^4 h$ term, so we
obtain $z^\mathrm{(II)} = 4$ and 
$\alpha^\mathrm{(II)}=\alpha_\mathrm{loc}^\mathrm{(II)} = 1/2$ 
in the early times (short scales) regime. Above the
crossover the $\nabla^2 h$ term governs the surface fluctuations and we find
$z^\mathrm{(I)}= 2/(1-\gamma)$ and 
$\alpha^\mathrm{(I)} = - (1-3\gamma)/[2(1-\gamma)]$ by simple 
power-counting. Therefore, for values $0 < \gamma \leq 1/3$ 
we have the desired situation of a crossover from a rough to a flat regime.
According to our scaling argument we expect scaling in the asymptotic regime 
with a roughness exponent $\alpha=\alpha^\mathrm{(II)}= 1/2$ 
and a dynamical exponent $z= 2/\gamma$, 
instead of the trivial values
$\alpha^\mathrm{(I)} = 0$ and $z^\mathrm{(I)}=0$ that would naively 
correspond to the $\nabla^2 h$ operator with conserved noise. 
A comparison with a numerical integration
of the model for a time-dependent coupling with exponents $0<\gamma<1/3$ 
shows an excellent agreement with this analysis (see Figure 2).
\begin{figure}
\includegraphics[width=0.40\textwidth]{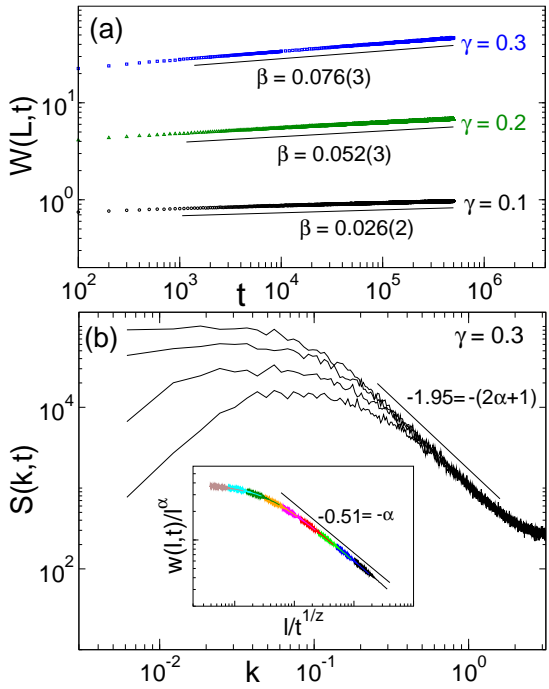}
\caption{(Color online) Crossover from rough to flat. 
Numerical integration of Eq. (\ref{eq:int}) 
with conserved noise in a system size $L=1024$. Results were averaged over 
200 realizations. (a) Global width  
for different $\gamma$. The growth exponent 
was found to fit with the theoretical value $\beta= \gamma/4$. 
The curves are vertically shifted for clarity.
(b) Power spectrum for 
$\gamma=0.3$ at times $10^{3}$, $10^{4}$, $10^{5}$, 
and $5\!\times\! 10^{5}$. It shows a roughness exponent 
$\alpha=0.48(3)$. The inset shows a data collapse of 
the local width data using 
$\alpha=0.51(2)$ and $z=6.63(4)$, which agrees 
with the scaling prediction $z=2/\gamma$ for $\gamma = 0.3$.\label{fig2}}
\end{figure}

\paragraph{Crossover to a damped regime.-}
Finally, we briefly discuss the case in which surface fluctuations become 
damped over a certain scale. This means that, 
for small deviations around the mean 
surface height, the terms $-h,\, -h^2,\, \dots$ are
to be included in the growth equation.
These terms break the $h \to h + c$ symmetry and therefore also break
scale-invariance.
Let us consider a growth model with the leading-order expansion 
${\cal G} = - \omega(t) h(x,t) + \omega_\mathrm{II}
\Phi_\mathrm{II}(x,t) + $ higher-order terms. In such a way that the operator
$\Phi_\mathrm{II}$ is less relevant than the $-h$ damping 
term, but it would lead to 
scale-invariant behavior in the absence of damping.
Following a scaling analysis similar to that of the
previous sections we find that for
$z^\mathrm{(II)} < q/\gamma$ the crossover to the damping regime will never 
take place. So, in all respects, the damping term is irrelevant and the
surface will exhibit the scaling behavior corresponding to the operator II.
On the contrary, if $z^\mathrm{(II)} > q/\gamma$ the crossover occurs when 
$\xi(t) \sim \ell_{\times}(t)$. The asymptotic regime in this case
does exhibit scaling but with the nontrivial exponents 
$\alpha=\alpha^\mathrm{(II)}$
 and $z = q/\gamma$, 
for the same reasons as in the previously discussed rough-to-flat case. 
Simulation results of the simple model (not shown),
$\partial_t h = -\omega(t) h + \nabla^{2} h + \eta$, 
are also in excellent agreement with this scaling analysis.

\paragraph{Conclusions.-} 
We have shown that the presence of time-dependent coupling in
non-equilibrium surface roughening has important 
implications in the scaling properties of the asymptotic regime. 
We have focused on local models that include the two most relevant 
terms separated by a crossover length scale 
growing in time, $\ell_{\times}(t)\sim t^{\gamma/q}$, which conspires 
with the correlation length of the system $\xi(t)\sim t^{1/z}$ to
give highly nontrivial scaling properties.
In the case of a crossover between two rough regimes, the surface 
may exhibit anomalous roughening, directly related to the value of  
$\gamma \neq 0$. On the other hand, in the case of a crossover from a 
rough to either a flat or damping regime, the dynamical crossover length 
$\ell_\times(t) \sim t^{\gamma/q}$ 
is the only relevant scale above the crossover and this immediately leads to 
an asymptotic rough regime with a dynamic exponent $z=q/\gamma$. A rough 
regime appears despite the most relevant term tends to flatten the surface. 
Remarkably, this is precisely the numerical result found in recent studies 
in spontaneous imbibition \cite{alava,dube,marc,laurila}, 
where the interface growth is described by 
Eq. (\ref{eq_imbi}). Our results show that these numerical results can be
understood in the wider context of kinetic roughening in systems with 
time-dependent couplings.

\acknowledgments 
This work is supported by the DGI of the Ministerio de Educaci{\'o}n y Ciencia 
(Spain) through Grant Nos. FIS2006-12253-C06-04 and -05.


\begin{thebibliography}{99}
\bibitem{note} An arbitrary small constant $c$ is
required to avoid the singularity at $t=0$, so that $\nu \sim \nu_0
(c+t)^{-\gamma}$. This has no significance for our analysis but it must be
taken into account in numerical simulations.
\bibitem{note2} Note that Eq.\ (\ref{eq:SPC}) describes the surface scaling
before saturation. For finite systems the steady-state
occurs at very long times, $t \sim L^{2/\gamma}$, where one recovers $S(k) \sim
k^{-4}$. We are interested here in the intermediate regime before saturation is reached,
in which both operators, $k^{-2}$ and $k^{-4}$, play a role at different scales.
\end{thebibliography}
\end{document}